# V2V Propagation Modeling with Imperfect RSSI Samples

Silvija Kokalj-Filipovic

Bin Cheng

Larry Greenstein

Marco Gruteser

WINLAB, Rutgers University
{skokalj,ljg,cb3974,marco}@winlab.rutgers.edu

*Abstract*— **We describe three in-field data collection efforts yielding a large database of RSSI values vs. time or distance from vehicles communicating with each other via DSRC. We show several data processing schemes we have devised to develop Vehicle-to-Vehicle (V2V) propagation models from such data. The database is limited in several important ways, not least, the presence of a high noise floor that limits the distance over which good modeling is feasible. Another is the presence of interference from multiple active transmitters. Our methodology makes it possible to obtain, despite these limitations, accurate models of median path loss vs. distance, shadow fading, and fast fading caused by multipath. We aim not to develop a new V2V model, but to show the methods enabling such a model to be obtained from in-field RSSI data.**

## I. INTRODUCTION

There has been considerable research to develop and calibrate a Physical (PHY) layer model for Vehicle to Vehicle (V2V) Dedicated Short Range Communications (DSRC) [1]. Such a model can be used to simulate large-scale V2V networks with statistical accuracy, thus allowing studies of V2V enabled driver safety applications. Developing the PHY model involves two important components. The first is the channel (propagation) model, which determines the power and delay of a transmitted signal at a receiver based on the distance between transmitter and receiver as well as certain environment-dependent parameters. The second component is a receiver model that determines when a receiver senses a clear or busy channel and when a packet is received successfully.

This paper concentrates on the channel model only, or, more accurately, on the modeling approach given field data. The particular challenge that we faced is to develop the channel model based on the received signal strength (RSSI) field data obtained from field experiments, which is an indirect and frequently corrupted indication of the channel quality, while a more accurate data would come from channel sounding methods, e.g., [2]. Yet, channel sounding is limited in scale, i.e. calibrating power and delay of the received signals requires expensive and sensitive equipment (per each participating vehicle), and cannot capture the multiple effects of an increasing number of simultaneous V2V transceivers mounted on the moving vehicles.

We note that a database consisting of RSSI samples is, by its very nature, not ideally suited to propagation modeling, for several reasons. One is that RSSI measures total power in the full system bandwidth (no frequency subdivision is done), leading to large noise floors. Another is that RSSI is reported only for packets sufficiently clean to be accurately received. Thus many packets are lost due to noise (especially at larger distances), and due to interference as well, in trials with many simultaneously transmitting terminals.

The aim of this paper is not to present a new V2V model, of which there are many, e.g., [3]-[5]. It is to showcase methods that can be used to extract useful models from imperfect data: the RSSI samples collected in typical environments and under dense traffic conditions, with vehicles frequently transmitting simultaneously. A particular contribution differentiating this work from other papers on RSSI-based models [6] is that we tackle RSSI impairments due to high density of transmitting vehicles. The paper describes our experimental approach, the problems with RSSI data, and the methods we have developed for working around these problems.

## II. PROBLEM DESCRIPTION

The data we use was collected by the Crash Avoidance Metrics Partnership (CAMP) Vehicle Safety Communications 3 (VSC3) Consortium, in partnership with the United States Department of Transportation (USDOT), as part of the V2V safety communications scalability activity of the CAMP's project *Interoperability Issues of Vehicle-to-Vehicle Based Safety Systems (V2V-Interoperability)*.

Note that RSSI measurements are made on OFDM transmissions with a 10MHz bandwidth centered near 5.9 GHz, in compliance with V2V DSRC standard IEEE802.11p [7], using Atheros 802.11p chips.

The field trials are marked by a certain scale of participating DSRC transceivers, ranging from 50 to 400 radios sending fixed size messages at a constant rate of $R$ Hz (messages per second). Almost every experiment of a certain scale was conducted for both 5 and 10 Hz. Note that the increasing message rate and increasing density of radios contribute to increasing interference which results in lost packets both due to collisions and lower SINR. These are the factors that need to be decoupled from the losses due to channel propagation issues (signal attenuation due to large distances or fading effects).

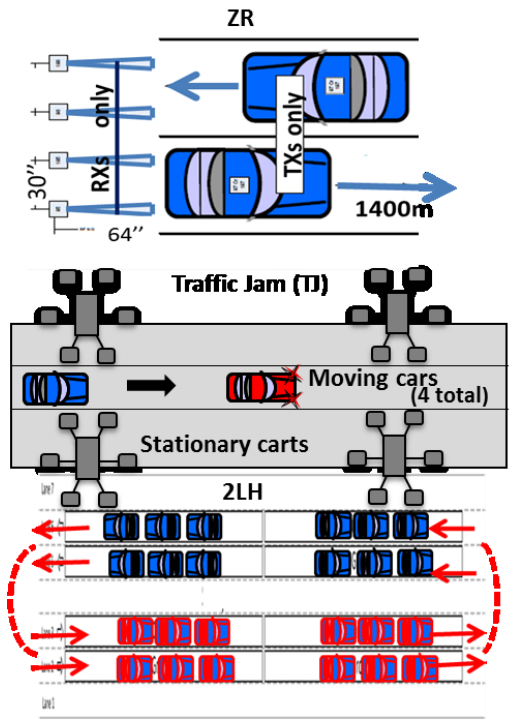

**Fig. 1:** Scenarios for the three trials. Top: Zero Reference (ZR); Center: Traffic Jam (TJ); Bottom: Two-Lane Highway (2LH).

Our work has benefited from several field trials, which were planned and conducted by the CAMP VSC3 V2V-Interoperability Team in cooperation with the USDOT, and we will focus here on three of them to demonstrate our methods. One is referred to as the *Zero Reference (ZR)* trial, which included only one transmitter at a time, mounted on a vehicle that traveled back and forth on a straight road 1200 m long (see Fig. 1, top). At one end of the road there were 4 pods with antennas spread 30 inches apart from each other. They logged RSSI values for packets sent from the transmitter and tagged by the distance from the transmitter. The ZR trial is the one with no interference, enabling us to separate the influences of interference and noise effects. While the absence of other vehicles changes the propagation environment from that with dense traffic (e.g., less scatter), this scenario helped us to study the impact of receiver noise on path loss modeling using RSSI data.

The second trial, another series of measurements on flat surface with line-of-site and open sky, helped us to study the effects of interference. This trial consisted of four cars moving in a single lane past 66 carts placed along a roadway (see Fig. 1, center). Each cart is made of a combination of steel and aluminum, and contains several OBEs (*On-Board Equipments*, i.e, DSRC transceivers); collectively, they emulated the interference from a typically dense spatial distribution of vehicles. The carts were stationary and distributed along 1200 m of roadway; and the total number of OBEs was 400. As the vehicles were passing along a dense deployment of 400 stationary transmitters, akin to what might occur in certain real-world traffic jam scenarios, we refer to it as *Traffic Jam (TJ)* trial. The absence of metallic surfaces--barring that of the four moving vehicles--resulted in reduced scatter, just as in the ZR trial. In both cases, the scatter plots of RSSI vs. distance take on the appearance of 2-ray propagation, as we will demonstrate and model.

The third trial modeled many vehicles moving under free-flow conditions along a 2-lane highway (*2LH*) in both directions (see Fig. 1, bottom). Here, the scatter was quite strong, leading to RSSI-distance scatter plots that are cloud-like, as in Fig. 2. The data from this scenario helped us to model the effects of scatter on path loss, to develop methods for modeling fast fading, and to further explore the mitigation of noise floor effects on modeling.

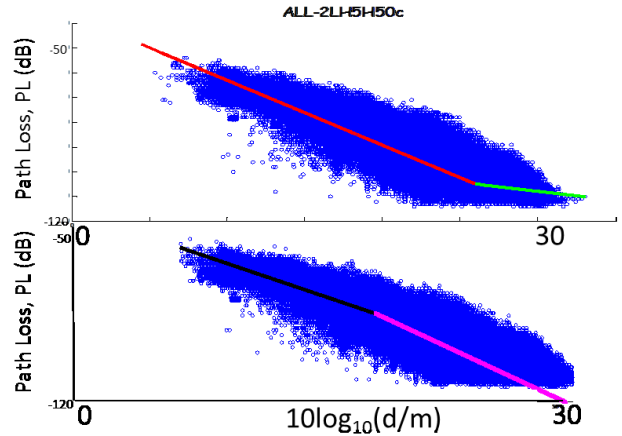

**Fig. 2:** Scatter plot of path loss vs. log-distance for 2LH data, showing 2-segment fits for the median path loss. Top: The second segment is made shallow by the noise floor. Bottom: For the same scatter plot, the noise floor effect is mitigated by additional processing.

### III. ISSUES WITH RSSI DATA

#### 3.1 Major Challenges

Here, we qualitatively characterize the issues in channel modeling with RSSI data and discuss solution approaches. We elaborate upon the latter in Section 4.

**Issue 1: Noise saturation of RSSI data at larger distances**
Noise saturation occurs when the received power is so small that noise dominates (low SNR), and most packets are lost. A clear example is given in Fig. 2, a scatter plot of recorded RSSI values and path distance in the 2LH trial. The trend with distance is downward, as expected, and a dense, fairly uniform cloud of points like this is well-fitted by a straight-line segment, or perhaps two segments with different slopes in different distance ranges. However, as seen, there is an RSSI level below which few if any points are recorded. This level corresponds to the receiver noise power, ~ -96 dBm in this case. Above some distance, most or all packets are dropped, creating a flat bottom edge of the scatter plot. Using least-squares estimation to create a 2-segment fit, we obtain the fit in Fig. 2, top. The shallow slope of the second segment is an artifact of the noise floor which, if somehow removed or worked around in the data processing, produces the result of Fig. 2, bottom. The steeper slope of the higher-distance segment is more consistent with realistic propagation [8][9]. We shall elaborate further on the methods used to achieve this improvement.

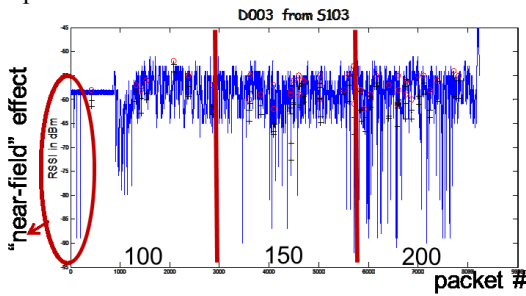

Fig. 3: Time plots for the 2LH trial, showing the effect of different numbers of active transmitters. The appearance of what seem like deep fades intensifies as the number increases from 100 to 150 to 200.

**Issue 2: False fading due to interference**
Fading (large and small scale) captures how the instantaneous signal level fluctuates over time, frequency, and space due to mobility and scattering. We present evidence that interference is included in the RSSI measurements, which takes the appearance of sudden fluctuations in the signal level. For large number of transmitters, these fluctuations resemble fast fading due to the law of large numbers (sums of multiple, random, time-shifted interference signals). Our analysis confirmed that under the same conditions except the number of active transmitters, these fluctuations are different. In a 2LH trial run with 200 vehicles on the ground, the test first included 100 active transmitters; then another 50 were added, and in the final third of the test all 200 vehicles were transmitting. The RSSI vs. time plot for a particular transmitter-receiver pair in this test is presented in Fig. 3, where the three parts of the trial are separated by vertical bars. Clearly, the fluctuation process intensifies with the number of active transmitters. Also, there are strong fluctuations at the beginning of the trial while the vehicles were static, which should produce no fast fading. We refer to this phenomenon as a "near-field effect", as we believe it is caused by vehicles being grouped together and inflicting stronger interference on each other. The same effect caused unusually high RSSI spread for small distances in a Traffic Jam trial run (Fig. 4, top). The appearance of deep fades for a static link in a stationary field test is also an artifact of interference (Fig. 4, center). A particularly difficult challenge was to decouple these interference effects from real fades in modeling small scale fading. For static nodes, it is easier to remove outliers (more about that in the next section) since we know that there should be no fast fading, and whatever appears to be a sudden deep fade (e.g., see Fig. 4, center) is an outlier, to be ignored. For mobile links, we used the ZR logs as the no-interference baseline to gauge the contribution of interference to RSSI fluctuations. The ZR RSSI fluctuations are quite shallow, as illustrated in Fig. 4 bottom.

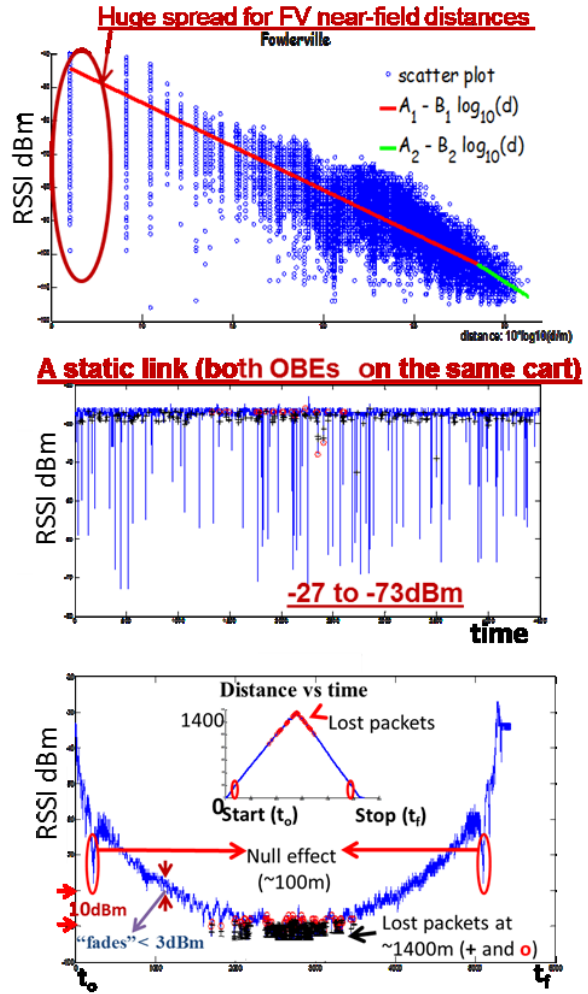

Fig. 4: Effects of interference on measured RSSI. Top: TJ data for RSSI vs. log-d, where interference accounts for the large spread at low distances. Center: TJ data for RSSI vs. time on a static link between OBEs mounted on the same cart. Bottom: ZR data for a mobile run (away and back), where the fluctuations for this interference-free scenario are shallow.

**Issue 3: Missing points due to interference**
Packets are lost not only at larger distances, where signal levels fall below noise; they can be lost when signal levels are good, due to interference. This is illustrated by Fig. 5, top, which shows, for the Traffic Jam trial, the number of recovered packets (red curve) and the number of packets lost (blue rectangles), both as functions of distance. The transmit (TX) power was 20 dBm. We see that severe losses occur even at distances where the signal should be strong compared to receiver noise; these losses are due to interference. At the larger distances, the packet losses are quite severe, indicating the effects of the receiver noise floor. Note the log scale of the distance causing an illusion of more packets at higher distances. In Fig. 5, bottom, we show results for a similar run in the ZR trial. One obvious change from Fig. 5, top is that there are no losses at the lower distances, a result of having no interference. The other change, related to the onset of packet losses due to noise, is discussed later.

### 3.2 Handling the Noise Floor (Censored Data)

One method for mitigating the influence of receiver noise is to increase the TX power. In the ZR trial leading to Fig. 5, bottom, the TX power was 26 dB, which is 6 dB above that used in the Traffic Jam trials leading to Fig. 5, top. As seen in the bottom plot, the onset of the noise saturation begins at higher distances and is a lot milder. Another example is given in Fig. 6, where data from the ZR trials are collected for two TX powers: 18 dBm (top) and 26 dBm (bottom). No 'bottoming out' of the scatter plot at large distances is evident on the bottom. The fitting of the straight-line segment is therefore better: the RMS variation about it is 1.6 dB, in contrast to 2.3 dB using a TX power of 18 dBm.

The second method is by "binning" the points over the distance range, and attempting to model the RSSI points within each bin as a Gauss-like distribution, based on the mode and histogram.

The expectation was that if we identify several bins wherein RSSI has a clean uni-modal distribution, we may infer the missing points based on assuming symmetry of the distribution. We refer to this binning approach as the *mode-fitting* method. Because of the lost points due to interference, the bin-based distributions of RSSI are not always uni-modal. Also, at and beyond some point in distance, the mode is likely to fall below the noise threshold, thus avoiding detection. However, if we can find several bins that have 'appropriate-looking' distributions and regard their RSSI modes as likely median values, we can fit a path loss function (e.g. linear or piece-wise linear) through these medians. This method was used to obtain the second segment in Fig. 2, bottom, which is steeper than the first segment, as expected [8][9].

### 3.3 Handling Interference

For **static links**, we model the median path loss and shadow fading based on a database obtained by calculating RSSI modes (most frequent RSSI values) for each transmitter-receiver pair.

Note that all RSSI readings for one transmitter-receiver pair have the same distance value (static OBEs). Since there is sufficient number of pairs for each distance value, we have sufficient statistics to model shadow fading and to obtain a complete path loss model.

For **mobile links**, we address the problem of interference-caused fluctuations by using only the ZR data set to model small-scale fading. With fast fading thus modeled, we can delineate the fast fading (multipath) model from the slow fading (shadowing) model, as explained in detail in the next section.

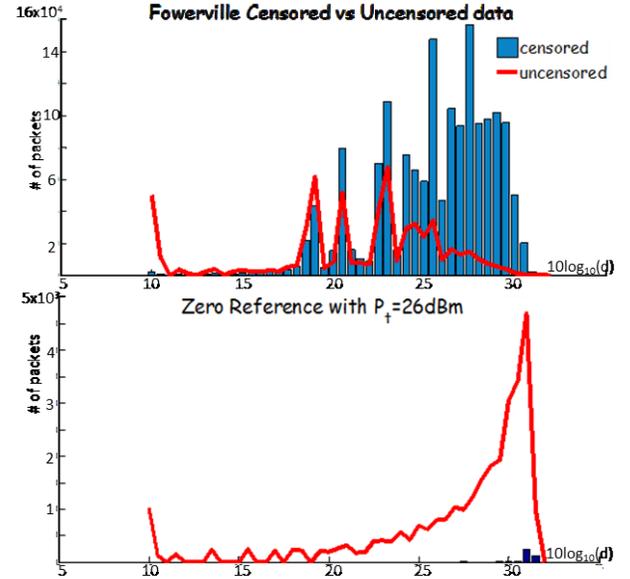

Fig. 5: Numbers of packets received vs. distance (10log(d)). Top: TJ logs for Pt = 20 dBm, where missing data are due not only to the noise floor, but also to interference. Bottom: ZR logs for Pt = 26 dBm, where there is no interference and the effect of the noise floor is diminished.

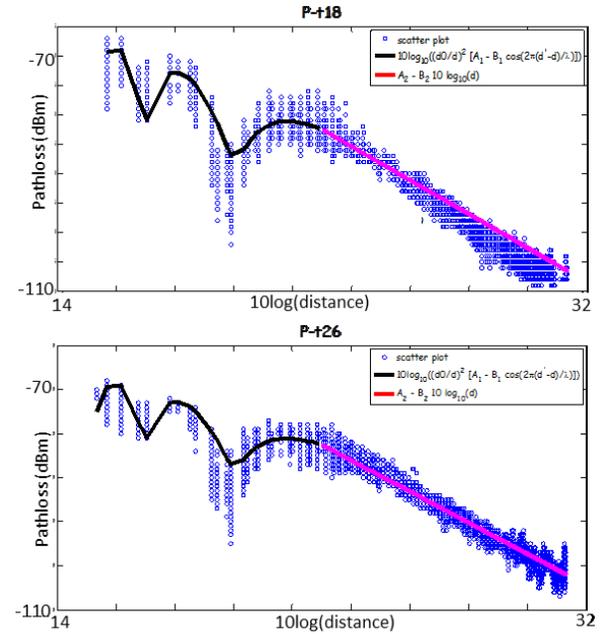

Fig. 6: Path loss vs. log-distance for ZR data, with Pt = 18 dBm (top) and 26 dBm (bottom); shows how higher power can be used to collect more data with complete statistics over the distance range of interest.

### IV. MODELING METHODOLOGY

### 4.1 Generic Path Loss

Following traditional methods for modeling wireless propagation, we write the *path gain* of a particular link (the ratio of received power ($p_r$) to transmit power ($p_t$)) as follows:

$$g = p_r/p_t = [g_{med}(d) F_{sh}] F_{mp} \quad (1)$$

where: the bracketed term is the locally averaged path gain for a particular receiver on the terrain at a distance $d$ from the transmitter; $g_{med}(d)$ is the median of this path gain taken over all transmitter-receiver links of length $d$; $F_{sh}$ is the deviation, for this particular link, of its locally averaged path gain from the median; and $F_{mp}$ is an additional deviation due to multipath fading. The term $F_{sh}$ varies in a seemingly random way from one transmitter-receiver link to another and is referred to as *shadow fading*; it changes slowly for a moving terminal, e.g., over distances on the order of tens to hundreds of meters, depending on the physical environment (urban, rural, suburban, etc) . The term $F_{mp}$ varies over space in a seemingly random way, a result of multipath scatter on the transmitter-receiver link; it changes rapidly, i.e., over travel distances of a wavelength or less. This term is scaled to have a local average value of one.

Some widely used conventions are to define the *path loss (PL)* as either the dB value of the bracketed term; or the dB value of $g_{med}$ alone; or the negative value of either of these. For our database, each PL value represents the dB $p_r$ in a given packet minus the dB $p_t$ (manufacturer-specific offset between the recorded RSSI value and the dB $p_r$ was taken into account in our processing by adjusting RSSI accordingly). Thus, we write the path loss as

$$PL = RSSI - 10 \log_{10} p_t$$
$$= G_{med} + 10 \log_{10}(F_{sh}) + 10 \log_{10}(F_{mp}), \quad (2)$$

where $G_{med}$ is the dB value of $g_{med}$. This formulation combines small-scale fading (through the term $F_{mp}$) and large-scale fading (through the term $F_{sh}$) although the RSSI measurement per packet does not (cannot) distinguish between the two. The need to treat them as separate terms is recognized in other published work, e.g., [10], which considers fast fading only but states statistical modeling for both kinds of fading as future work. We will show one way to separate them in our processing

## 4.2 Median Path Loss

In our various trials, we have discerned two distinct patterns of RSSI vs. $d$ at low-to-moderate distances (i.e., $d$ up to 200 m):
- from the Traffic Jam and Zero Reference trials, we see the effects of 2-ray propagation, wherein the dominant rays consist of a direct (line-of-sight) ray and a ray reflected from the road (see Fig. 6);
- from the 2LH trial, we see a 'cloud' of points suggestive of a lot of multipath scatter rather than dominance by two rays (see Fig. 2).

We have thus devised two approaches to modeling $G_{med}$ in this $d$-range. For higher distances, all trials suggest the use of a straight-line segment of RSSI vs. log-$d$, where the linear segment is contiguous with the segment for $G_{med}$ in the lower $d$-range.

**Two-ray Modeling:** For a 2-ray mode of propagation, with the transmitter and receiver antennas at a common height $h$, $p_r$ can be described in the approximate form (see Appendix):

$$p_r = p_t(d_0/d)^2[A_1 - B_1 \cos(2\pi(d'-d)/\lambda)], \quad (3)$$

where $d_0$ is a reference distance (we chose $d_0 = 10$ m) and

$$d' = [d^2 + h^2]^{1/2} \quad (4).$$

The constants $A_1$ and $B_1$ have a particular form in 2-ray theory [11], but we know that this picture can be modified, in the scenarios we studied, by other objects and irregularities in the road. Consequently, we leave these constants general and use the actual data to determine them, using least square estimate. Thus, RSSI values in the database are de-logged to get values of $p_r$; $A_1$ and $B_1$ are chosen to get a least-squares fit for $p_r/p_t$ vs. $d$; and the result is converted to dB to obtain PL vs. $d$. This was done for the Zero Reference and Traffic Jam data for $d$ out to some breakpoint distance, $d_{br}$, typically, less than 200 m (See Fig. 6 for the Zero Reference case).

**Linear-segment Modeling:** For $d > d_{br}$ in the Traffic Jam and Zero Reference data, our data show a cloud of points suitable to modeling by a straight line of RSSI vs. log-$d$. Thus, for the second distance region in all cases, we propose a path loss function of the form

$$PL = A_2 - B_2 (10 \log_{10}(d/d_0)); \quad d > d_{br} \quad (5)$$

The selection of $A_2$ and $B_2$ is again based on least square estimate fitting to data, subject to the constraint that the second segment matches the first segment at the breakpoint distance, $d_{br}$, as in Fig. 6. It is important to note that the choice of $d_{br}$ was based on trial-and-error, by doing the fittings for many candidate breakpoint distances and choosing the one leading to the lowest RMS deviation of data points about the fitted curves.

For scenarios such as 2LH, Fig. 2, the scatter plots of RSSI vs. $d$ indicate linear segments in both distance regions or, in some cases, the use of just one segment. As indicated in Section 3.1, the second segment derived from data can have an artificially shallow slope, due to the dominance of noise saturation in the distance region for that segment (Fig. 2, top). Using the method of binning and mode fitting described in Section 3.2, an improvement to the slope of this segment can be achieved to first order, as shown in Fig. 2, bottom.

## 4.3 Slow (Shadow) Fading

There is a large body of evidence, e.g., [11]-[13] supporting the notion that the shadow fading term $F_{sh}$, in our path loss formula is Gaussian. A simple representation for $F_{sh}$ is

$$F_{sh} = \sigma u \quad (6)$$

where $u$ is a zero-mean, unit variance Gaussian random variable, and $\sigma$ is the standard deviation of the scatter of RSSI points about the median fit. We note that this result can apply to the entire distance range being fitted, under the assumption that the shadow fading is constant over distance; or $\sigma$ can vary over distance, and this can be quantified via binning, i.e., computing the mean and standard deviation of RSSI values within uniformly-spaced distance bins. Note, however, that the RSSI data includes the variation about the median caused by both shadowing and multipath. Therefore, the standard deviation computed from data is actually

$$\sigma = [\sigma_{sh}^2 + \sigma_{mp}^2]^{1/2}, \quad (7)$$

where the standard deviations for the two kinds of fading seem to be inextricably coupled. We now show that, in fact, they can be separated by ascertaining the fast fading distribution.

### 4.4 Fast (Multipath) Fading

Portions of the database offer long sequences of RSSI values with no packet losses (close-up in Fig. 7). These sequences, with the RSSI values de-logged to yield sequences of $p_r$, can be processed to estimate the underlying fading pdf. This is helped by the fact that, for the packet rates, wavelength and vehicle speeds involved, adjacent RSSI samples were assumed to be independent.

The process used was as follows:
**(a)** Select long blocks of successive RSSI measurements, with few or no packet losses, for the same transmitter-receiver link. These "RSSI signatures" are the ones processed to determine fast-fade distributions.
**(b)** For each block, de-log the RSSI values to get a sample sequence of $p_r$.

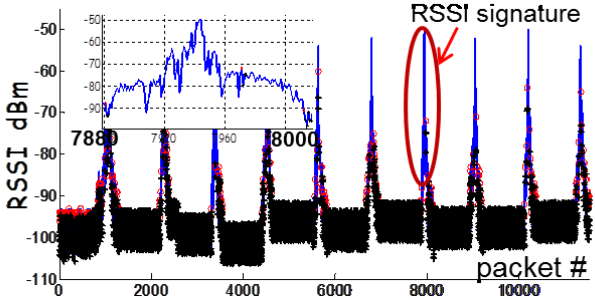

Fig. 7: 2LH logs of RSSI vs. time, where black areas denote unreceived packets. By extracting "signature" data blocks with few losses (the selected signature is blown up in the inset), processing can be used to estimate fast fading distributions.

**(c)** Use a P-packet-wide sliding window to smooth out the fast fluctuations (P>>1), thereby obtaining an approximate local average of $p_r$.
**(d)** Use that average to normalize the P power samples, repeat over the next P packets, and continue this way over the entire block. The resulting block of normalized samples should have an autocorrelation function (ACF) that indicates low sample-to-sample correlations (see Fig. 8). If this is not the case, a different value of P is selected.
**(e)** With the full set of samples, try different **pdf**s (Nakagami, Gaussian), using both visual sighting and Kolmogorov-Smirnov testing [14].
This procedure is repeated over other long blocks of clean RSSI samples. An example of cumulative distribution functions (CDF), comparing theoretical to empirical CDFs, is shown in Fig. 8.

Once a satisfactory ACF and a good CDF fit are achieved, it is easy to determine, from any empirical sequence of power samples, the standard deviation of the corresponding dB values. This is $\sigma_{mp}$, the standard deviation of $F_{mp}$ in our path loss formula.

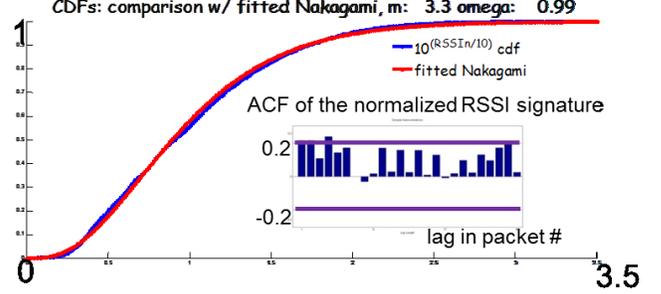

Fig. 8: Empirical and Nakagami CDFs for a block of normalized linear-power samples extracted from 2LH data. The ACF of the samples (inset) shows that they are basically uncorrelated.

With $\sigma$ thus determined from the path loss fitting, and $\sigma_{mp}$ computed from the fast fading procedure above, it is easy to determine $\sigma_{sh}$, the standard deviation of the shadow fading, according to equation (7).

### V. EVALUATION AND DISCUSSION

### 5.1 Model Evaluation

We evaluate how accurately our modeling methodology reproduces field data against a conventional log-normal fit. We implement in ns-3 simulator the model obtained by applying our methodology to one set of TJ data, and then run the simulation with the same node movement as in the field trial. For the conventional log-normal fit, we use the linear model of RSSI v.s. log-d (i.e., eqn. (5)) to fit the median pathloss into the field data, and a Gaussian model for the shadow fading. The fast fading is not modeled in this fit. The metrics we use to compare the simulation and the field trial results are Packet Error Rate (PER) and the 95th percentile Inter-Packet Gap (IPG). The PER is defined as the percentage of lost packets at a receiver from one particular transmitter. The IPG is the measurement of the elapsed time between successively received packets from a particular transmitter. We calibrate ns-3 simulator by setting the same PHY and MAC layer parameters obtained from the field trial setting (TX power 20 dBm, packet rate 10 Hz, TX rate 6Mbps, bandwidth 10 MHz, packet size approximately 300 bytes etc.). The propagation parameters used in our model are: $d_{br} = 400m$, $A_1 = 7.31e^{-7}$, $B_1 = 3.79e^{-7}$, $\sigma_1 = 5.25$ (2-ray segment), and $A_2 = 18.58$, $B_2 = 4.30$, $\sigma_2 = 5.03$ (linear fit). The parameters for the log-normal model are $A = -31.99$, $B = 2.45$, $\sigma = 5.39$.

*Evaluation Results:* PER and IPG for two mobile OBEs are calculated for both the simulation and the field trial results. The chosen OBEs are mounted on one vehicle moving past all the carts. Both PER and IPG are organized into 40 meters bins. For each bin, we also calculate the average value for the two OBEs. The absolute error with respect to field trial PER or IPG at each distance bin is computed and plotted. Fig. 9 shows the absolute error of the PER for both modeling methods. We observe that the error for the proposed model is generally less than 5% and that it performs better than the log-normal model, especially in the short range region. One reason is that our model successfully captures the two-ray feature of the field trial data, which is not represented in the log-normal model. Our model also demonstrates better performance in the 95th percentile IPG, as depicted in Fig. 10. This is because our model is able to reproduce the interference more accurately.

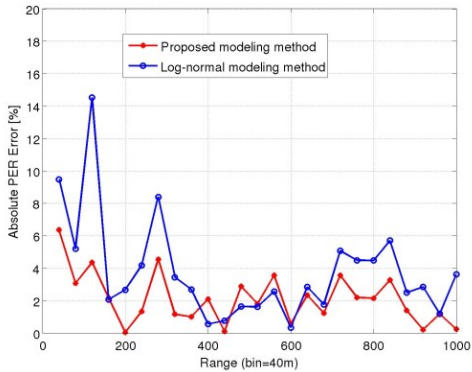

**Fig. 9 The absolute error between the PER from the simulations and that from the field trial.**

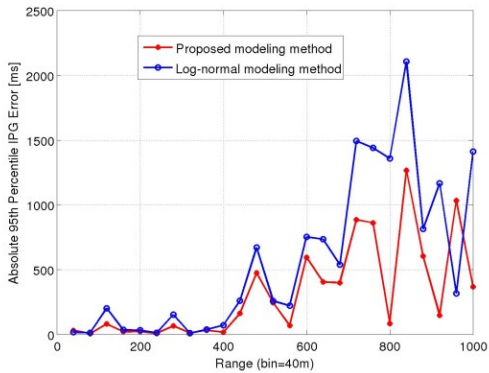

**Fig. 10 The absolute error between the 95th percentile IPG from the simulations and that from the field trial.**

### 5.2 Interference and Noise Effects

We have been largely successful in decoupling the effects of interference on channel quality from the effects caused by fading. Perhaps the greatest challenge moving forward will be to mitigate the effects of receiver noise in wide (V2V-specific) bandwidths, especially at larger distances. The mode-fitting method described here has produced some good results. An alternative approach along these lines is reported in [15], where propagation parameter estimation is based on Expectation Maximization (EM) [16][17] in the presence of censored (lost) data. Here, the assumption is that the per-bin distributions of RSSI points are not only symmetrical, but Gaussian. This is a plausible premise, given the many reports of Gauss-distributed path loss data. Moreover, the reported method also makes use of the (measurable) numbers of received and lost data values (as in Fig. 5). The use of this method in data studies like the present one is worth consideration.

However, a major obstacle is that censored data is due not only to the noise floor, but also to interference, as illustrated in Fig. 5, top, while [15] considers the noise floor only (as in Fig. 5, bottom). The difference between cases exemplified by the bottom and top plots in Fig. 5 is this: In the former case, the RSSI values of dropped packets are all below some fixed RSSI threshold (noise floor) while, in the latter case, RSSI values of dropped packets are randomly distributed within the range of measurable RSSI values. Our ongoing work is focused on decoupling those two types of packet losses so that parameter estimation in the presence of censored data can be sufficiently accurate regardless of the transmitter density.

Fig. 11 illustrates the way interference compromises the Gaussian assumption. In each graph, the straight line corresponds to the quantile-quantile (QQ) plot for a Gaussian distribution, and the other curve is the plot for actual RSSI data. Future work should include efforts to quantify the deviation from a Gaussian distribution as a function of the number of simultaneous (i.e., interfering) transmissions, and to mitigate its impact on the propagation modeling.

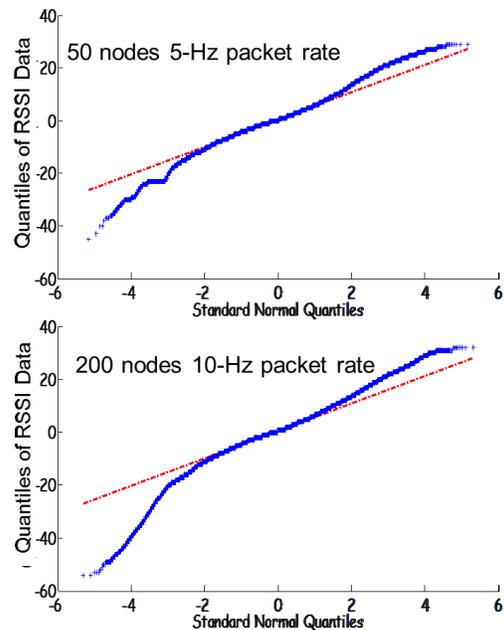

**Fig. 11: QQ plots for 2LH data, comparing empirical distribution of RSSIs with Gaussian distribution. The greater deviation from Gaussian in the bottom plot illustrates how higher transmitter density and increased packet rate compromise the Gaussian distribution.**

## VI. CONCLUSION

Despite the fact that RSSI measurements in the field, with many radios transmitting at once, cannot be as clean or as flexible in measurement bandwidth as one would desire, they are still a valuable resource for propagation modeling. The main virtue of such data is that they are obtained at a low cost compared to the use of channel sounding methods, hence allowing large-scale experiments illustrative of real-world V2V network phenomena, including fading and interference due to high density of mobile vehicles. We have shown several methods, with examples based on field data, which permit models to be extracted. Properly applied, these methods can yield accurate and usable results, for specified scenarios, of the three critical model components: Median path loss vs. distance (including environments with either 2-ray dominance or high scatter); statistics of slow (shadow) fading; and statistics of fast (multipath) fading. Future work will focus on further decoupling interference and noise effects and on mitigating the impact of both on data reductions.

### APPENDIX: FITTING 2-RAY MODEL TO MEASURED DATA

Consider 2 antennas at height $h$ above flat ground and separated by distance $d$. Ideally, there will be 2 rays arriving at the receiver antenna (Rx) from the transmitter antenna (Tx): a line-of-sight ray, arriving over a path of length $d$, and a reflected ray, arriving from a ground bounce at mid-path of length

$$d' = 2[(d/2)^2 + h^2]^{1/2} = [d^2 + 4h^2]^{1/2} . \quad (A\text{-}1)$$

The electric field amplitude of the first ray is $a(d_0/d)$, where $a$ is a complex scale factor representing the field magnitude at reference distance $d_0$; and the wave amplitude of the second ray is $b(d_0/d')$, where the complex scale factor $b$ includes the reflection coefficient $c_r$ of the ground bounce. Both $a$ and $b$ depend, as well, on the elevation patterns of the transmit and receive antennas and the transmit power, $p_t$. Under typical conditions of the ground and the wave polarization, $c_r \approx -1$, but we need make no such assumptions. We will, however, assume the negative polarity of the second ray, and write the received sum of the 2 rays, for Tx-Rx separation $d$, as follows:

$$E(d) = [a(d_0/d) \exp(-j2\pi d/\lambda) - b(d_0/d') \exp(-j2\pi d'/\lambda)]. \quad (A\text{-}2)$$

Following convention, we regard the received power as the squared magnitude of $E(d)$. Thus,

$$p_r(d) = |E(d)|^2 = A' - B' \cos(2\pi(d' - d)/\lambda), \quad (A\text{-}3)$$

where

$$A' = [a(d_0/d)]^2 + [b(d_0/d')]^2 \text{ and } B' = 2ab(d_0^2/dd'). \quad (A\text{-}4)$$

We need not conform to this idealized result, as we know that the actual conditions are a bit different. However, it seems plausible that, for some scenarios (e.g., ZR), the 2-ray situation roughly prevails, so we retain the form of (A-3).

*Argument of the cosine term.* It is obvious from (A-3) that $p_r$ has its lowest values wherever the argument of the cosine term is a multiple of $2\pi$. Thus, the dips seen in the data should occur when $d' - d = n\lambda; \quad n = 1, 2, 3, \ldots$

Given the definition of $d'$ in (A-1), it is easy to solve for the values of $d$ where this condition is met:

$$d_n = (4h^2 - (n\lambda)^2)/2n\lambda; \quad n = 1, 2, 3, \ldots \quad (A\text{-}5)$$

The largest distance for which a dip occurs corresponds to the case $n = 1$. In our database, the wavelength $\lambda$ is 0.0512 m. Inserting these values into (A-5), we find that the final dip occurs at $d = 100$ m when $h = 1.6$ m, which seems to be consistent with data.

*Fitting A' and B' to data.* In (A-4), every term has a denominator of $d^2$, $d'^2$ or $dd'$. Also, for $d > 6h$ (~10 m), (A-1) shows that we can reasonably approximate $d'$ by d. The result is that, for $d > 10$ m, we can reasonably model $p_r(d)$ as

$$p_r(d) = p_t (d_0/d)^2 [A_1 - B_1 \cos(2\pi(d' - d)/\lambda)], \quad (A\text{-}6)$$

where $A_1$ and $B_1$ are $A'$ and $B'$, with $p_t (d_0/d)^2$ factored out.

### ACKNOWLEDGMENT

We thank the CAMP VSC3 V2V-Interoperability Team and the USDOT for planning and executing the requisite tests and providing the raw data measurements and sponsoring the work that enabled this analysis. The CAMP VSC3 Consortium consists of the Ford Motor Company, General Motors LLC., Honda R&D Americas, Inc., Hyundai-Kia America Technical Center, Inc., Mercedes-Benz Research and Development North America, Inc., Nissan Technical Center North America, Inc., Toyota Motor Engineering & Manufacturing North America, Inc. and Volkswagen Group of America.